\documentclass[a4paper,fleqn,usenatbib]{mnras}

\usepackage{newtxtext,newtxmath}

\usepackage[T1]{fontenc}
\usepackage{ae,aecompl}

\usepackage{graphicx}	
\usepackage{amsmath}	
\usepackage{amssymb}	
\usepackage{graphicx}	
\usepackage{amsmath}	
\usepackage{amssymb}	
\usepackage[shortlabels]{enumitem}      
\usepackage{booktabs}   
\usepackage[nice]{nicefrac}  

\newcommand{\blue}[1]{{\color{blue}#1}}
\newcommand{\orange}[1]{{#1}}

\graphicspath{{Figures}}
\DeclareGraphicsExtensions{.png,.pdf,.eps}

\title[Predicting DMAF with cGANs]{A black box for dark sector physics: Predicting dark matter annihilation feedback with conditional GANs}

\author[F. List et al.]{
Florian List$^{1}$\thanks{E-mail: flis0155@uni.sydney.edu.au},
Ishaan Bhat$^{2}$, and 
Geraint F. Lewis$^{1}$
\\
$^{1}$Sydney Institute for Astronomy, School of Physics, A28, The University of Sydney, NSW 2006, Australia \\
$^{2}$UMC Utrecht, Image Sciences Institute, 3508 GA Utrecht, The Netherlands
}

\date{Accepted XXX. Received YYY; in original form ZZZ}

\pubyear{2019}

\begin{document}
\label{firstpage}
\pagerange{\pageref{firstpage}--\pageref{lastpage}}
\maketitle

\begin{abstract}
Traditionally, incorporating additional physics into existing cosmological simulations requires re-running the cosmological simulation code, which can be computationally expensive. We show that conditional Generative Adversarial Networks (cGANs) can be harnessed to predict how changing the underlying physics alters the simulation results. To illustrate this, we train a cGAN to learn the impact of dark matter annihilation feedback (DMAF) on the gas density distribution. The predicted gas density slices are visually difficult to distinguish from their real brethren and the peak counts differ by less than $10$ per cent for all test samples (\orange{the average deviation is $< 3$ per cent}). Finally, we invert the problem and show that cGANs are capable of endowing smooth density distributions with realistic substructure. The cGAN does however have difficulty generating new knots as well as creating/eliminating bubble-like structures. We conclude that trained cGANs can be an effective approach to provide mock samples of cosmological simulations incorporating DMAF physics from existing samples of standard cosmological simulations of the evolution of cosmic structure.
\end{abstract}



\section{Introduction}
Since the advent of early $N$-body simulations in the 1960's where the gravitational interaction between up to $N \approx 100$ dark matter macro-particles could be described, cosmological simulation codes have developed from purely gravitational solvers to multiphysics programs that involve baryonic effects on different scales, such as gas cooling, star formation, feedback from supernovae and supermassive black holes, etc. (see e.g. \orange{\citealt{Vogelsberger2014, Vogelsberger2019, Schaye2015, Springel2017}}). \orange{Starting from random primordial density perturbations, these simulations model the collapse of matter into cosmic structure over the course of billions of years. This process does not result in randomly distributed galaxies, but rather in the formation of dense clusters, long filaments, thin sheets, and empty voids in between the other constituents -- a structure that is referred to as the \emph{cosmic web} \citep{Bond1996} and that can be observed in our Universe (e.g. \citealt{Dietrich2012}).} The complexity of \orange{cosmological} simulations necessitates large computational clusters and extensive computational times  since a set of non-linear partial differential equations (commonly the Euler equations) needs to be solved for billions of collisional particles, in addition to the evaluation of gravitational forces, feedback and the application of subgrid physics.  Computationally cheaper alternatives to particle- or grid-based cosmological simulations are semi-analytic models (\citealt{White1991, Kauffmann1993, Somerville2008} amongst others) and mathematical formalisms for galaxy formation -- most prominently the Press--Schechter formalism \citep{press1974formation} and extensions thereof \citep{bond1991excursion, Sheth1999}. 
\par Recently, machine learning and in particular deep convolutional neural networks, found their way into cosmology and have produced remarkable results in areas such as the evolution of disc galaxies \citep{Forbes2019}, the construction of galaxy catalogues \citep{Khan2019}, linking haloes and galaxies to dark matter \citep{KodiRamanah2019, Zhang2019}, the prediction of galactic properties from dark matter haloes \citep{Kamdar2016}, galaxy model fitting \citep{Aragon-Calvo2019}, modelling the disruption of subhaloes due to baryonic feedback \citep{Nadler2018}, the classification of the cosmic web \citep{Aragon-Calvo2019a}, the estimation of cosmological parameters from the dark matter distribution \citep{Ravanbakhsh2016} and from weak gravitational lensing convergence maps \citep{Ribli2019}, and map generation for weak lensing \citep{Mustafa2019} and dark matter \citep{Jeffrey2019}, to name but a few. 
\par A promising avenue is to use neural networks for modelling structure formation, complementary to the aforementioned traditional methods. 
A specific type of deep generative model, \orange{Generative Adversarial Network (GAN)}, has been employed for creating realistic samples of the cosmic web in \citet{Rodriguez2018}. The GAN is trained on output from $N$-body simulations and learns to generate new random dark matter density slices. An extension to the creation of three-dimensional samples using a multi-scale approach and a comprehensive analysis of the results are presented in \citet{Perraudin2019}. \citet{Troster2019} use GANs \orange{and Variational Auto-Encoders (VAEs)} for predicting gas pressure distributions from the dark matter density and find that maps of the Sunyaev--Zel'dovich effect created from the GAN-generated mock distributions match those from hydrodynamical simulations.
\orange{\citet{He2019} show that deep neural networks can be utilised for predicting non-linear structure formation from an initial displacement field. They train a neural network on displacement fields generated from $N$-body simulations and achieve a higher accuracy than second-order perturbation theory, which is the traditional fast method for the prediction of structure growth in the non-linear regime.} Whereas an expensive cosmological simulation is needed to evolve each initial displacement field in time, the evaluation of trained neural networks on a new set of initial conditions typically only takes a few seconds or less. The same rigidness applies to changing parameters, varying physical models, and introducing additional physics in cosmological simulations, all of which make new simulation runs necessary.
\par In this proof-of-concept paper, we investigate the question as to whether neural networks are able to forecast how a \emph{change in the underlying physics}, which possibly affects the entire history of the simulated universe, impacts the distribution of a physical quantity at a certain point in time, say at redshift $z = 0$. 
The structure of this paper is the following: in Section \ref{sec:cGANs}, 
we introduce \orange{conditional GANs (cGANs)} and show how they can be exploited for predicting output from cosmological simulations. In Section \ref{sec:DMAF}, the physics of \orange{dark matter annihilation feedback (DMAF) from} WIMP-like \orange{dark matter} particles, which smooths the cosmic gas web and quenches the formation of galaxies and stars, is briefly explained. Section \ref{sec:sims} is dedicated to the set-up of the cosmological simulations and the data creation for the neural network. In Section \ref{sec:results}, we present our results and assess the quality of the neural network output visually as well as by means of power spectral density and peak counts. In Section \ref{sec:reverse}, we discuss the arguably more challenging task of reversing the direction of inference to DMAF $\to$ no DMAF, which requires predicting a more complicated density distribution from a simpler one. We conclude this work in Section \ref{sec:conclusions}. 

\section{Predicting structure with cGANs}
\label{sec:cGANs}
\orange{
\subsection{The need for fast emulators of the cosmic web}
In the coming decades, large galaxy surveys will tighten the current constraints for cosmological parameters and probe dark sector models at unprecedented precision across the entire electromagnetic spectrum. This quest relies on the gathering and evaluation of gargantuan amounts of data, not only for ground-based instruments such as the LSST \citep[\emph{Large Synoptic Survey Telescope},][]{abell2009lsst} that is expected to generate about 20 TB of raw data per day, but also increasingly for space telescopes such as Euclid \citep{refregier2010euclid} and WFIRST \citep[\emph{Wide-Field Infrared Survey Telescope},][]{spergel2015wide}, with a daily data volume of $\sim$ 100 GB (Euclid) and $\gtrsim$ 1 TB (WFIRST). The surge in observational data volume goes hand in hand with the need for ever larger cosmological simulations for preparing and testing the data processing and analysis pipelines years before the first observations are made. The Euclid Flagship simulation \citep{potter2017pkdgrav3}, for instance, tracked the evolution of 2 trillion dark matter $N$-body particles over billions of years of cosmic history to the present age,
and the resulting galaxy catalogue comprises over 2 billion galaxies. Although the scaling of numerical algorithms for $N$-body simulations has improved from direct summation (of order $\mathcal{O}(N^2)$) over the Particle-Mesh method \citep{hockney1988computer} and the Barnes--Hut method \citep{Barnes1986} (both $\mathcal{O}(N \log N)$) to the Fast Multipole Method \citep{Greengard1987} and Multigrid solvers \citep{Brandt1977} (both $\mathcal{O}(N)$), the amount of CPU (and of late often also GPU) time and memory required for such simulations is enormous. Furthermore, varying cosmological parameters \citep{Kacprzak2016}, incorporating extensions to the standard $\Lambda$ cold dark matter ($\Lambda$CDM) cosmological model \citep{Baldi2012}, or comparing different baryonic feedback models \citep{Duffy2010} requires running a suite of simulations for exploring the desired parameter space. In order to save computational time, fast emulators have been developed which interpolate in the parameter space \citep{Lawrence2010} or use probabilistic approaches \citep{Lin2015}.
\par Recently, \citet{Rodriguez2018} and \citet{Perraudin2019} introduced GANs as an alternative fast method for creating new artificial samples of the cosmic web. As a complementary approach, we propose to train GANs on multiple cosmological simulations with different parameters or physics to learn the resulting variations of the cosmic web. 
\par In the long run, we envision  replacing the need for meticulously exhausting parameter spaces with simulation suites by trained neural networks that map the cosmic web generated by one (or a few) reference simulations to that corresponding to a different set of parameters (which might differ from those used for training the cGAN).
More specifically for the case of DMAF considered in this work, a cGAN trained on a suite of simulations spanning a range of DMAF strengths could readily add the effects of DMAF with an arbitrary strength in a fraction of the time required for a new cosmological simulation. In practice, it would be interesting to examine whether a cGAN that has been trained on small simulation boxes is able to make reliable predictions when evaluated on a larger volume, possibly with a higher particle resolution. Then, a toolbox of pre-trained cGANs for various effects could be applied to state-of-the-art simulations in order to produce artificial simulation suites covering ample parameter spaces and physics models.
\par Clearly, further investigating the generalisation properties of GANs in view of different simulation box sizes, resolutions, and parameter ranges is necessary, and conditioning a cGAN on an input image \emph{and} a parameter such as the DMAF strength is a challenging task. Encouraging results on the generalization properties of cGANs has have been reported by \citet[][hereafter \blue{pix2pix}]{Isola2017}, who show that a cGAN trained on $256 \times 256$ images performs well on $512 \times 512$ images for a map $\leftrightarrow$ areal photo task, and in the context of cosmology by \citet{KodiRamanah2019}, who present a GAN that is able to predict haloes for arbitrary box sizes.
\par As a first step towards the vision of versatile cGAN emulators for cosmological simulations, we train a cGAN on two-dimensional density slices for a fixed DMAF strength and analyse the predictions of the cGAN in this paper.
}

\subsection{Tweaking physics as an image-to-image translation task}
Typically, the raw output of a cosmological simulation at each output time consists of the particle positions $\mathbf{p}_N \in \mathbb{R}^{3 N}$, together with the physical quantities at each particle (or cell, for adaptive mesh refinement codes), e.g. $\mathbf{q}_N \in \mathbb{R}^{d N}$, where $N$ denotes the number of particles for which the respective quantity is defined and $d \in \mathbb{N}$ is the dimension of the quantity. For sufficiently large particle numbers, these values can be viewed as tracers of an underlying continuous field $\mathbf{q}$, which can be approximated on a regular grid by interpolating the values of $\mathbf{q}_N$ onto the grid coordinates. For a uniform grid made up of $(M \times M \times M) \in (\mathbb{N} \times \mathbb{N} \times \mathbb{N})$ points, one thus obtains values $\mathbf{q}_G \in \mathbb{R}^{M \times M \times M \times d}$ on the regular grid. Assuming the values of $\mathbf{q}$ are normalised to the unit interval $[0, 1]$, $\mathbf{q}_G$ can be interpreted as a three-dimensional image with $d$ channels. Naturally, if slices through the simulation box are considered, the resulting image $\mathbf{q}_G \in \mathbb{R}^{M \times M \times d}$ is two-dimensional.
\par From this viewpoint, the prediction of the change in a physical quantity from one simulation to another can be regarded as an image-to-image translation problem. This is where neural networks come into play since they are known for performing well in image-to-image translation tasks \orange{(\blue{pix2pix}; \citealt{Choi2018})}. During the training process, the neural network is iteratively fed input data and tries to produce an output that is similar to the ground truth. After each iteration, its free parameters are updated in order to generate images that resemble the ground truth more closely in the next iterations. 
\par The notion of ``similarity'' between images requires further specification.
The simplest approach is to model the individual pixels of the image as being independent of each other and to minimise a pixelwise loss function such as the Euclidean distance \orange{ $d_{L^2}(\mathbf{a}, \mathbf{b}) = \sqrt{\sum_m (a_m - b_m)^2}$} or the $L^1$ distance \orange{$d_{L^1}(\mathbf{a}, \mathbf{b}) = \sum_m |a_m - b_m|$} between the predicted image and the ground truth. However, this approach is prone to produce blurry output because pixelwise minima result from an averaging over the space of possible values \citep{Pathak2016, 10.1007/978-3-319-46487-9_40}. Moreover, in most real world applications, neighbouring pixels are correlated and form connected patches that vary smoothly. An elegant way to utilise multipixel information for quantifying the quality of a predicted output image is to take a second neural network, the so-called discriminator, whose task it is to distinguish images created by the generating neural network from the real ones. This is the basic idea of a cGAN, which has been shown to achieve impressive results in image-to-image translation (\blue{pix2pix}) and which we use for our purposes. 

\subsection{Generative Adversarial Networks}
Generative Adversarial Networks (GANs) were first introduced in \citet{Goodfellow2014} and have been dubbed ``the coolest idea in machine learning in the last twenty years'' \citep{LeCun2016}. The idea of a GAN consists in letting two players, a \emph{generator} $G$ and a \emph{discriminator} $D$, play a min-max game against each other in order to achieve convergence towards a Nash equilibrium \citep{nash1950equilibrium}. The generator is a map that takes a random noise vector $\mathbf{z}$ (usually sampled from a standard normal or uniform distribution) as an input and produces an output image, i.e. $G: \mathbf{z} \mapsto G(\mathbf{z})$. The discriminator takes an input $\mathbf{u}$ and assigns a probability that $\mathbf{u}$ is drawn from the ground truth distribution, $D: \mathbf{u} \mapsto D(\mathbf{u}) \in [0, 1]$. The input for $D$ is either a ground truth sample, that is $\mathbf{u} = \mathbf{y}$, or stems from the generator, $\mathbf{u} = G(\mathbf{z})$. While $D$ tries to discern real and fake input, $G$ aims at producing output that is indistinguishable from the ground truth samples. The functions $D$ and $G$ are differentiable with respect to the network parameters such that a stochastic gradient descent method (most notably the Adam optimiser \citep{Kingma2014}) can be applied for updating the weights of the neural network. The optimal mapping $G^*$ is given as the solution of the saddle point problem
\begin{equation}
\label{eq:GAN}
\begin{aligned}
    G^* = \mathrm{arg} \min_G \max_D \Big(&\mathbb{E}_{\mathbf{y} \sim p(\mathbf{y})} \big[ \log(D(\mathbf{y})) \big] \\
    + &\mathbb{E}_{\mathbf{z} \sim p(\mathbf{z})} \big[ \log(1 - D(G(\mathbf{z}))) \big]\Big).
\end{aligned}
\end{equation}
Assuming a perfect discriminator, the above expression boils down to the Jensen--Shannon divergence between the GAN-generated distribution and the ground truth output distribution \citep{Goodfellow2014}. Alternative loss functions that rely on different metrics have been developed such as Wasserstein GANs \citep{Arjovsky2017}, which are based on the Wasserstein distance (also known as earth mover's distance).
Although GANs are highly heuristic in nature and deriving rigorous convergence results is hard given the complexity of the networks and the data they seek to create, some theoretical work has been done, for example addressing the question of how well the trained output distribution approximates the underlying ground truth distribution \citep{Arora2017, Arora2017a}. 

\subsection{The conditional variant}
\begin{figure}
  \centering
  \noindent
  \resizebox{\columnwidth}{!}{
  \includegraphics{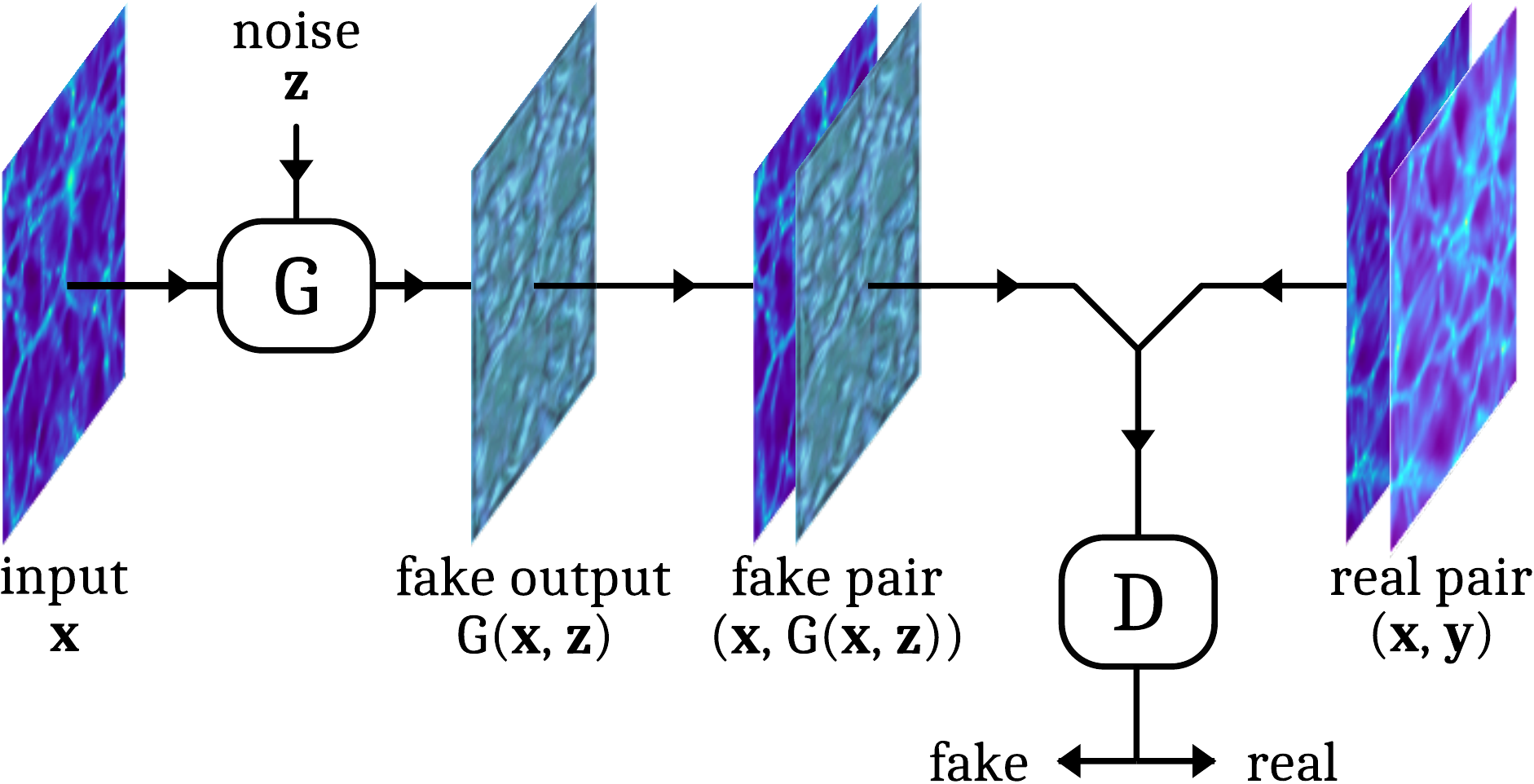}
  }
  \caption{Schematic representation of a cGAN: given the input $\mathbf{x}$ and random noise $\mathbf{z}$, the generator $G$ creates a fake output $G(\mathbf{x}, \mathbf{z})$. The task of the discriminator $D$ is to learn whether an input pair $(\mathbf{x}, \mathbf{u})$, consisting of the input image $\mathbf{x}$ and either a real output $\mathbf{u} = \mathbf{y}$ or a fake output $\mathbf{u} = G(\mathbf{x}, \mathbf{z})$, is real or fake.}
  \label{fig:cGAN_sketch}
\end{figure}
For unconditional GANs as described above, the distribution of the output generated by $G$ is expected to converge towards the distribution of the ground truth $p(\mathbf{y})$. Even for a fully converged GAN, however, it is impossible to determine \emph{a priori} which particular image will be sampled next from $p(\mathbf{y})$ as this is purely determined by the random noise vector $\mathbf{z}$. Since our goal is to relate input images from one simulation to output images from another, we need to couple the generated output \orange{images} to the original input \orange{images}. The conditional variant of a GAN, cGAN \citep{Mirza2014}, naturally lends itself to this end. In a cGAN, the input image $\mathbf{x}$ is shown to both the generator and the discriminator, i.e. $G = G(\mathbf{x}, \mathbf{z})$, $D = D(\mathbf{x}, \mathbf{u})$; in other words, $G$ and $D$ are \emph{conditioned} on the input data $\mathbf{x}$. The saddle point problem that describes the resulting Nash equilibrium then takes the form
\begin{equation}
\label{eq:cGAN}
\begin{aligned}
     G^*  = \mathrm{arg} \min_G \max_D \Big(&\mathbb{E}_{\mathbf{x} \sim p(\mathbf{x}), \mathbf{y} \sim p(\mathbf{y})} \big[ \log(D(\mathbf{x}, \mathbf{y})) \big] \\
    + &\mathbb{E}_{\mathbf{x} \sim p(\mathbf{x}), \mathbf{z} \sim p(\mathbf{z})} \big[ \log(1 - D(\mathbf{x}, G(\mathbf{x}, \mathbf{z}))) \big]\Big).
\end{aligned}
\end{equation}
While a cGAN learns to distinguish fake from real images without the need to specify a comparison criterion, it has proven useful to combine the cGAN loss with a pixelwise loss function (\citealt{Pathak2016}, \blue{pix2pix}). As the $L^1$ error is known for introducing less blur than the $L^2$ error, we modify the min-max problem to 
\begin{equation}
\label{eq:cGAN_L1}
\begin{aligned}
    G^* = \mathrm{arg} \min_G \max_D \Big(&\mathbb{E}_{\mathbf{x} \sim p(\mathbf{x}), \mathbf{y} \sim p(\mathbf{y})} \big[ \log(D(\mathbf{x}, \mathbf{y})) \big] \\
    + &\mathbb{E}_{\mathbf{x} \sim p(\mathbf{x}), \mathbf{z} \sim p(\mathbf{z})} \big[ \log(1 - D(\mathbf{x}, G(\mathbf{x}, \mathbf{z}))) \big] \\
    + \lambda \, &\mathbb{E}_{\mathbf{x} \sim p(\mathbf{x}), \mathbf{y} \sim p(\mathbf{y}), \mathbf{z} \sim p(\mathbf{z})} \big[\|G(\mathbf{x}, \mathbf{z}) - \mathbf{y} \|_1 \big]\Big),
\end{aligned}
\end{equation}
where $\lambda$ controls the contribution of the pixelwise $L^1$ error. Herein, we choose $\lambda = 100$.
\par The principle of a cGAN is schematically illustrated in Figure \ref{fig:cGAN_sketch} for the case of images of the gas density distribution from simulations where dark matter interacts only gravitationally (input $\mathbf{x}$) or actively injects energy created by DMAF into the gas (real output $\mathbf{y}$), see Section \ref{sec:DMAF}. 
The generator $G$ creates a fake gas density distribution in an attempt to reproduce the effects of DMAF. The task of the discriminator is to map input-output pairs with fake output to $0$ and pairs with real output to $1$. For a perfect generator that produces the underlying ground truth output distribution, the expectation value of the discriminator output will be $0.5$ since it is not able to discern real and fake output.

\subsection{Neural network architecture}
\begin{figure*}
  \centering
  \noindent
  \resizebox{\textwidth}{!}{
  \includegraphics{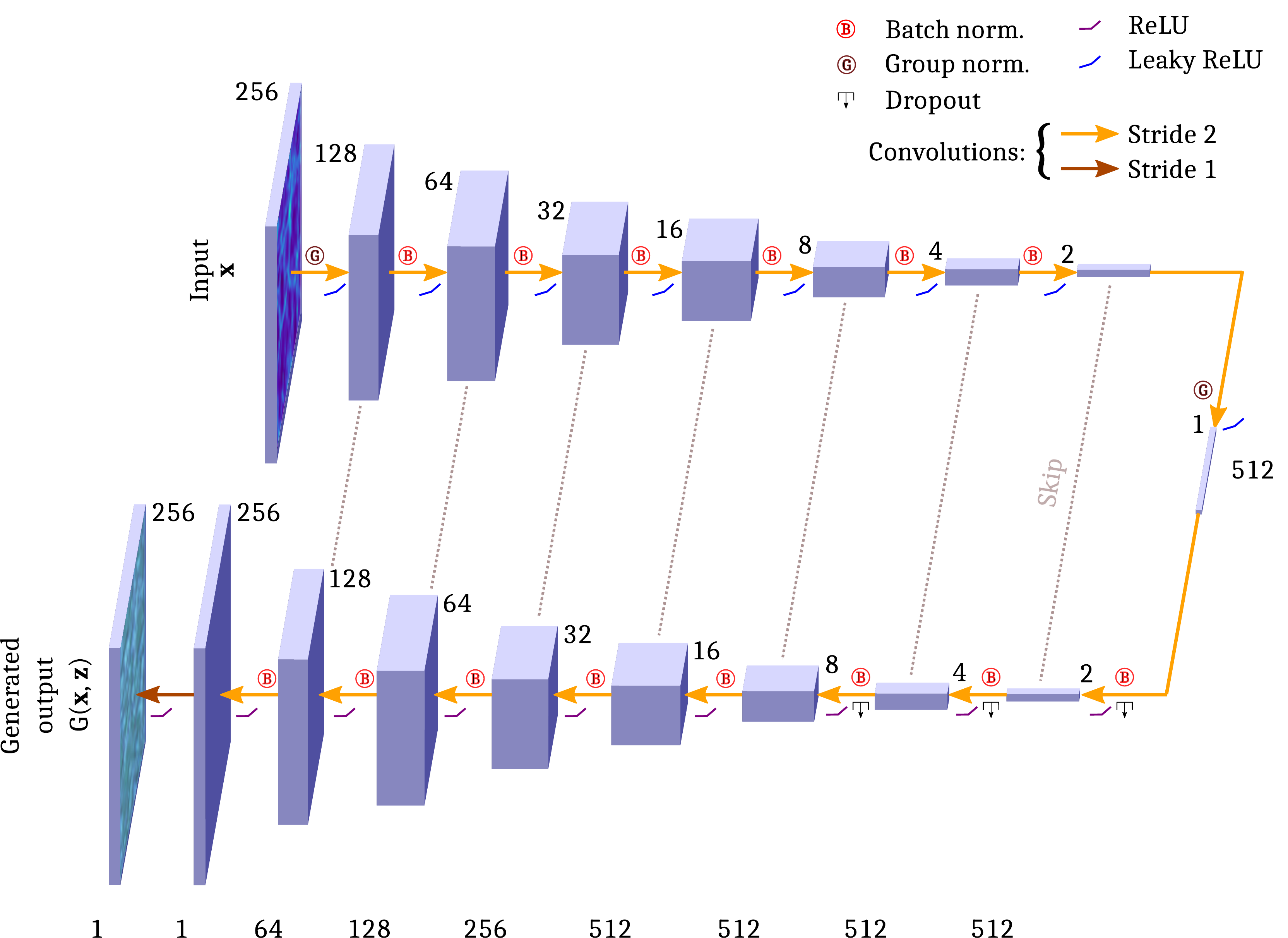}
  }
  \caption{Network architecture of the \emph{generator} (U-Net). The cubes depict the layers and the numbers indicate spatial resolution as well as number of channels. The dimensions of the tensors change due to the convolutions, which are followed by normalisation, dropout, and activation operations. The encoding path and decoding path are linked via skip connections, which are realised by concatenating tensors from both paths with each other.}
  \label{fig:Generator}
\end{figure*}

\begin{figure}
  \centering
  \noindent
  \resizebox{\columnwidth}{!}{
  \includegraphics{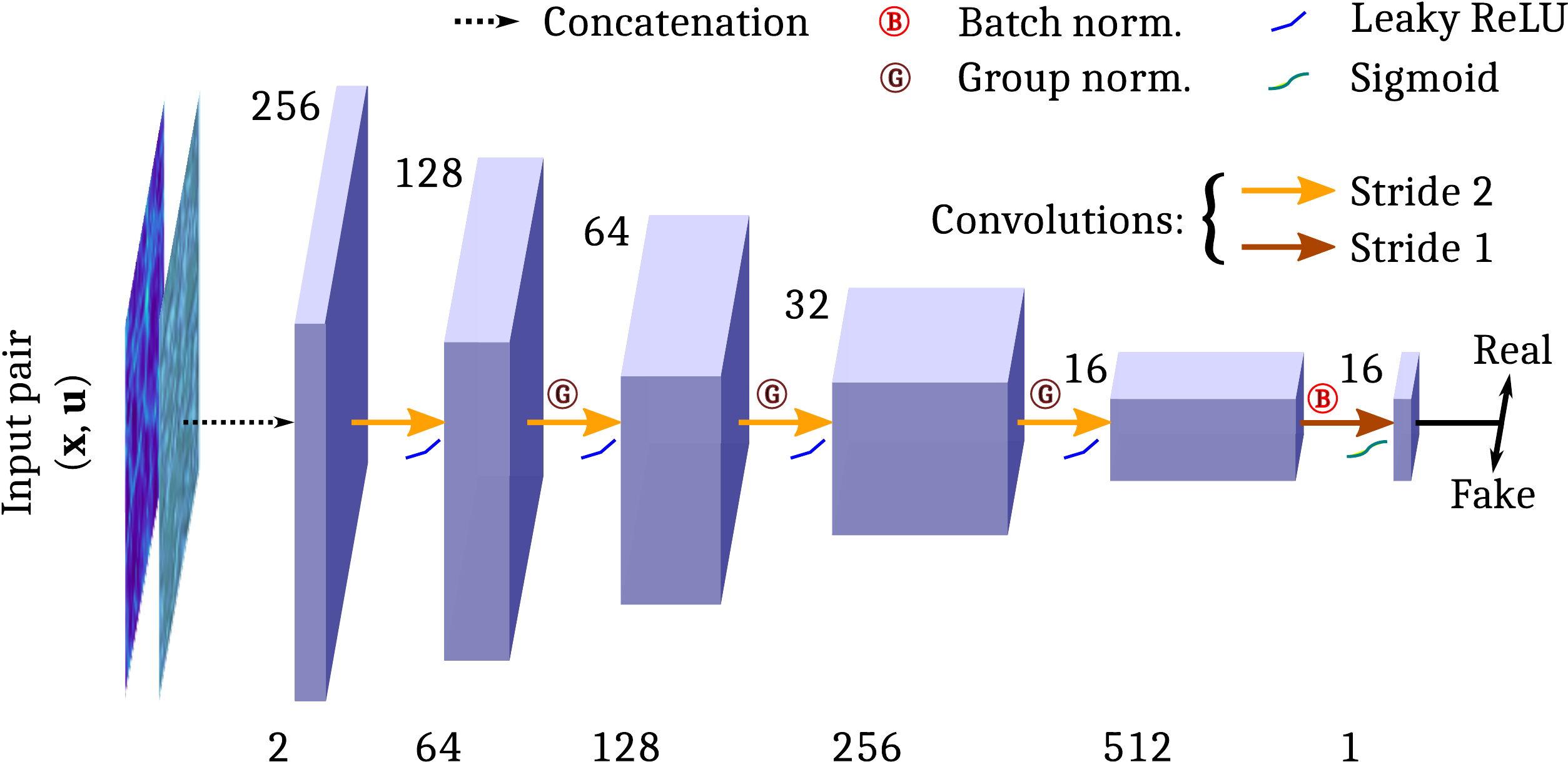}
  }
  \caption{Network architecture of the \emph{discriminator} (PatchGAN). The discriminator receives a real or fake input pair $(\mathbf{x}, \mathbf{u})$ as an input and attempts to distinguish fake from real pairs. The size of the receptive field, based on which the discriminator renders its judgement, is determined by the properties of the convolutional layers.}
  \label{fig:discriminator}
\end{figure}
In constructing our cGAN, we follow closely the architecture proposed by \blue{pix2pix} which has been shown to be effective in various applications (see the examples in said paper). For the generator, we use a U-Net \citep{Ronneberger2015} that consists of a downwards encoding path and an upwards decoding path, which are connected by skip connections at each layer (see Figure \ref{fig:Generator}). This is similar to \orange{auto-encoder} networks \citep{Hinton2006}, but thanks to the additional skip connections between the encoding and decoding path, the localisation of features is superior to classic \orange{auto-encoders}. Additionally, the skip connections improve the flow of gradients. In the encoding path, the spatial resolution decreases at an ever increasing number of channels (up to 512), whereas in the decoding path, the initial spatial resolution is gradually retrieved while the number of channels decreases. This bottleneck architecture with respect to spatial resolution encourages the neural network to learn compressed representations of hidden features which, combined with the spatial information from the encoding path that evades the bottleneck via the skip connections, leads to accurately reconstructed images with the desired properties at a high degree of locality. The layers of the generator are connected by convolutions that mix information from various pixels, batch normalisations and group normalisations for normalising the input (see \citealt{Ioffe2015, Wu2018}), dropouts with probability $p = 0.5$ for preventing overfitting \citep{Hinton2012, srivastava2014dropout}, and activation functions (ReLU and leaky ReLU with slope 0.2) that mimic the activation of a neuron in the human brain. The last operation is a convolution with stride $1 \times 1$ that results in one output channel (for the logarithmic gas density at each pixel), followed by a ReLU layer for \orange{enforcing non-negative output values. Often, a tanh layer is used as the final layer (and the images are scaled to $[-1, 1]$ instead of $[0, 1]$) but we achieved better results with a final ReLU layer.} 
For further details on the components of the neural network, we refer the reader to the standard literature, \orange{e.g. \citet{geron2017hands}}.
\par For the discriminator, we take a PatchGAN as proposed in \blue{pix2pix}. In contrast to a pixelwise loss function, the receptive field of a PatchGAN comprises patches consisting of several pixels, which enables the discriminator to gauge the quality of the generated images on a multipixel scale. This leads to sharp images since artificial blur in the patches will be penalised. Since the authors of \blue{pix2pix} report that increasing the receptive field of the discriminator to the entire image size does not give any improvement, we opt for a receptive field size of $94 \times 94$ pixels, generated by the discriminator architecture sketched in Figure \ref{fig:discriminator}. The final convolutional layer maps the tensor to a single output channel, and we apply a sigmoid activation function subsequently.
\par In our experiments, it proved beneficial to replace batch normalisation layers in the discriminator by group normalisation layers with group size 32 \citep{Wu2018}. Moreover, we use a group normalisation layer in the bottleneck of the generator in order not to zero out the previous layer for batch size 1 (see the Errata in \blue{pix2pix}). We use a convolution kernel of $5 \times 5$ pixels for the generator and $4 \times 4$ pixels for the discriminator since this gave the best results in our experiments. 
\par Altogether, the combined network consisting of generator and discriminator has 87,774,173 trainable parameters. 

\subsection{Implementation and training process}
The implementation of the neural network is done using Tensorflow \citep{Abadi2016}. As suggested by \citet{Goodfellow2014}, the generator minimises $- \log(D(\mathbf{x}, G(\mathbf{x}, \mathbf{z})))$ instead of $\log(1 - D(\mathbf{x}, G(\mathbf{x}, \mathbf{z})))$ in order to avoid shallow gradients. The generator noise $\mathbf{z}$ is modelled by the various dropout layers in the decoder path and we do not feed any additional noise. 
We minimise the losses for generator and discriminator in an alternating manner. 
\par We train our neural network for $600$ epochs on $300$ training images (see Section \ref{subsec:data_creation}) using an Adam optimiser with learning rate $2 \times 10^{-4}$, momentum parameters $\beta_1 = 0.5$ and $\beta_2 = 0.999$, and $\varepsilon = 10^{-8}$ for numerical stability. In our tests, using larger batch sizes did not improve the training, for which reason we \orange{report our results for} batch size 1. The training is carried out on an Nvidia K80 GPU card on the supercomputer Raijin, which is located in Canberra and is part of Australia's National Computing Infrastructure (NCI). 
\par During the inference phase when the trained cGAN is tested on unseen images, batch normalisation is done using moving averages of moments calculated during the training phase. The dropout layers remain in the neural network, leading to a small stochasticity in the predicted output. We repeated the calculation of the quantities in Section \ref{sec:results} multiple times and checked that the fluctuations of the statistics are sufficiently small. 

\section{Dark matter annihilation feedback}
\label{sec:DMAF}
Accurate measurements from the Planck satellite \citep{PlanckCollaboration2018} and other instruments show that $\sim 84.3$ per cent of the matter in the Universe consists of dark matter, which does not emit any electromagnetic radiation. However, the nature of dark matter is still one of the biggest unsolved mysteries in cosmology. The mass range of possible dark matter candidates spans more than 70 orders of magnitudes and while the exclusion bounds for the parameter ranges of promising candidates are gradually becoming tighter (e.g. \citealt{Leane2018}), no confirmed detection of dark matter has been reported to date.
\par One of the leading candidates for dark matter is weakly interacting massive particles (WIMPs) that were in a thermal equilibrium with the baryonic matter in the early Universe. A possible mechanism for this equilibrium to be maintained is dark matter pair annihilation into standard model (SM) particles. The classic scenario for explaining the current dark matter abundance is the so-called freeze-out which occurred when the expansion of the Universe became rapid enough for the WIMPs to fall out of equilibrium. From the observed relic dark matter density today, the velocity cross-section of WIMP annihilation into SM particles can be estimated to be $\langle \sigma v \rangle \approx 3 \times 10^{-26} \ \text{cm}^3 \ \text{s}^{-1}$ (see \citealt{Steigman2012} for a detailed derivation).
\par While $\gamma$-ray and antiparticle detectors such as Fermi and AMS-02, respectively, are seeking high-energy particles originating from annihilation of WIMPs \citep{Hoof2018, Cholis2019, Cuoco2019, Leane2019}, a complementary approach is to investigate the imprint of energy from DMAF onto the large-scale structure. The specific rate of DMAF energy $u$ absorbed by the surrounding gas is given by
\begin{equation}
    \label{eq:DMAF}
    \frac{du}{dt} = \frac{\langle \sigma v \rangle}{m_\chi} \frac{\rho_\chi^2}{\rho_g} c^2,
\end{equation}
where $m_\chi$ is the dark matter particle mass, $\rho_\chi$ and $\rho_g$ are the mass densities of dark matter and gas, respectively, and $c$ is the vacuum speed of light.
For simplicity, we neglect a boost factor here \citep{Bergstrom1999} and make the assumption that the entire energy produced by dark matter annihilation is deposited into the gas (see \citealt{Slatyer2009, Galli2013, Madhavacheril2014} for specific absorption fractions for different annihilation channels and redshifts).
DMAF has been shown to affect structure formation, suppressing the formation of stars \citep{Ascasibar2007, Schon2015, Schon2017} and modifying the shape of galaxies \citep{Natarajan2008, Wechakama2011}. 
\par A self-consistent method for integrating DMAF into cosmological simulations is presented in \citet{Iwanus2017}. As shown in \citet{Iwanus2019} where an in-depth analysis of the results using this method for different dark matter masses is carried out, a major effect of DMAF is to erase small-scale structure in the gas density distribution and to smear the filamentary structure. 
\par Predicting slices of the washed out gas density field arising from a cosmological simulation with DMAF given density slices from simulations without DMAF provides a suitably challenging test case for exemplarily showcasing the performance of cGANs, given the non-linearity of the interplay between hydrodynamics, gravitation, and energy injection from DMAF.

\section{Simulations}
\label{sec:sims}
\subsection{Simulation set-up}
For our simulations, we use the simulation code \textsc{Gizmo} \citep{GIZMO}, which is a progeny of the \textsc{Gadget} series \citep{Gadget1, Gadget2} and comes with modern hybrid methods that combine the strengths of particle- and grid-based codes. For the work herein, we use the meshless finite volume method. \orange{In order to emphasise the effects of DMAF, we do not adopt any baryonic physics models for processes such as radiative cooling or star formation. In particular, the reference simulation without DMAF is adiabatic. Since heat generated by baryonic feedback from Active Galactic Nuclei (AGN) and stars affects the gas in a similar way as DMAF (albeit on possibly different scales), we expect that cGANs are useful for predicting these processes as well.} The DMAF is implemented using the donor-based method introduced in \citet{List2019} with solid-angle based weights (see the reference for more details). \orange{This method is based on a discretised form of Equation \eqref{eq:DMAF}, whereby the dark matter density is evaluated at each $N$-body dark matter particle at each time-step. Then, the energy generation is calculated and the resulting energy is deposited into the surrounding gas particles. The injection mechanism is determined by weights that can be chosen to account for the deposition length of the annihilation channel at hand. For solid-angle based weights, the energy is localised to the $N_\text{receiver}$ nearest $N$-body gas particles and distributed amongst them in an (approximately) isotropic way. Here, we choose $N_\text{receiver} = 40$}. 
\orange{Since the dark matter mass loss due to DMAF is negligible after the WIMP freeze-out that occurred fractions of a second after the Big Bang, we keep the dark matter particle masses constant in our simulations.} 
\par The simulation box is a cube of side length \orange{$100 \ {h}^{-1} \ \text{Mpc}$} in a co-moving frame, containing $512^3$ gas and dark matter particles, respectively.  Here, ${h}$ is the dimensionless Hubble parameter, defined by $H_0 = 100 \ h \ \text{km} \ \text{s}^{-1} \ \text{Mpc}^{-1}$. \orange{The particle masses are $1.48 \times 10^8 \ \text{M}_\odot$ (gas) and $7.95 \times 10^8 \ \text{M}_\odot$ (dark matter).} The initial conditions at our starting redshift $z = 100$ are created with the tool \textsc{N-GenIC} \citep{NGenic} which relies on the Zel'dovich approximation \citep{zel1970gravitational}. The cosmological parameters are taken from \citet{Ade2016} and are given by $\Omega_m = 0.3089$, $\Omega_b = 0.0486$, $\Omega_\Lambda = 0.6911$, and $H_0 = 67.74 \ \text{km} \ \text{s}^{-1} \ \text{Mpc}^{-1}$. At the faces of the box, periodic boundary conditions are imposed. Physical quantities are reconstructed using $N_\text{ngb} = 40$ neighbour particles. The gravitational smoothing length is taken to be \orange{$9.77 \ {h}^{-1} \ \text{kpc} $ and the minimum allowed kernel width for gas particles is $2.44 \ {h}^{-1} \ \text{kpc}$}. For the dark matter particle, we assume a thermal relic velocity cross-section of $\langle \sigma v \rangle \orange{=} 3 \times 10^{-26} \ \text{cm}^3 \ \text{s}^{-1}$ and \orange{take a particle mass of $m_\chi = 10 \ \text{MeV}$.} \orange{Note that a dark matter particle this light is at odds with the current lower bound of $m_\chi \gtrsim 20 \ \text{GeV}$ for s-wave $2 \to 2$ annihilation of a thermal WIMP into visible states \citep{Leane2018}. However, the resulting high energy production leaves a strong imprint on the gas density distribution, which aids to visually assess the quality of the cGAN-generated samples and facilitates the analysis of the power spectral density and peak counts that are heavily modified by DMAF. These modifications need to be reproduced by the cGAN.} The dark matter particle mass only determines the strength of DMAF and is irrelevant for the reference simulation without DMAF.

\subsection{Data creation for the neural network}
\label{subsec:data_creation}
From the gas density fields at redshift $z = 0$, we extract $128$ uniformly spaced gas density slices in each coordinate plane using the package \textsc{yt} \citep{Turk2011} and thus obtain $384$ slices in total; each with a resolution of $256 \times 256$ pixels. Using two-dimensional data instead of the three-dimensional density has two advantages: firstly, it massively speeds up the training of the neural network since each image contains $256$ times less pixels as compared to the full 3D case and secondly, a simulation in a three-dimensional box provides a large number of two-dimensional slices that can be used for training whereas the generation of a substantial amount of three-dimensional training data would require running more simulations or extracting small subvolumes. The simulated density slices bear resemblance to the well-known slices of the cosmic web from large-scale galaxy surveys such as 2dF \citep[\emph{Two-degree-Field Galaxy Redshift Survey},][]{Colless2001} and SDSS \citep[\emph{Sloan Digital Sky
Survey},][]{Blanton2017}. The density distributions in neighbouring slices (\orange{$\sim 780  \ h^{-1} \ \text{kpc}$} apart from each other) share many similarities but each slice still introduces new features that need to be learned by the neural network. In order to further increase the variability of the data, we randomly flip each slice horizontally and/or vertically with a respective probability of $0.5$. For obtaining sufficient contrast, we take the decimal logarithm of the density at each pixel. Finally, we scale the logarithmic densities linearly to the range $[0, 1]$, where the lower and upper limits correspond to $-2$ and $4 \ \log_{10}(\text{M}_\odot \ \text{kpc}^{-3})$ and values outside this range are mapped to the interval boundaries. More complex choices of transfer functions are proposed in \citet{Rodriguez2018, Perraudin2019}. We shuffle the resulting images, randomly choose $300$ as training data, and leave the remaining $84$ as test data. 

\section{Results}
\label{sec:results}
\begin{figure*}
  \centering
  \noindent
  \resizebox{\textwidth}{!}{
  \includegraphics{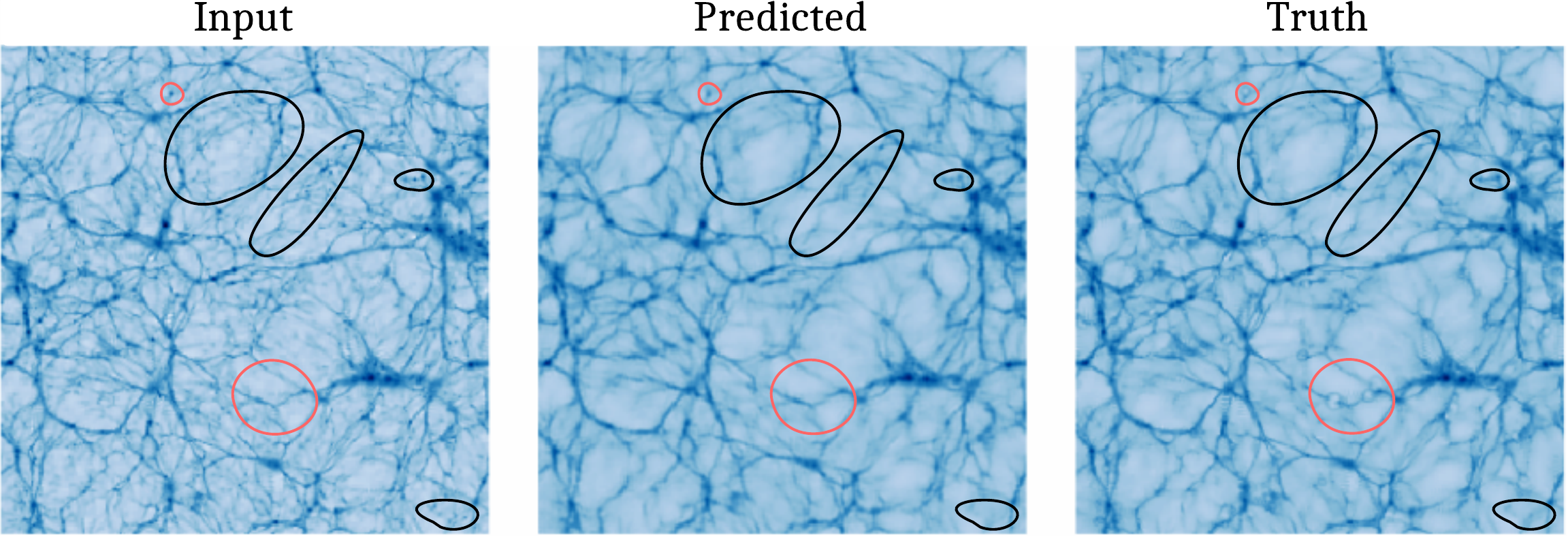}
  }
  \caption{\emph{No DMAF $\to$ DMAF}: predicted gas density \orange{distribution for an image from the test set}. The left \orange{image is the input without DMAF}, the \orange{central image depicts the prediction by the cGAN}, and the \orange{right image has been generated from the cosmological simulation with DMAF}. Some salient features are highlighted \orange{in black and red/grey (for positive and negative examples, respectively)}: much of the substructure has been washed out and the robust filaments that remain resemble the ground truth (two large \orange{black} loops). Many knots have been smeared (upper small \orange{black} loop) or erased (\orange{black} loop in the lower right corner). The formation of bubbles in the DMAF simulation due to large amounts of energy injected has not been mimicked by the neural network -- bubbles neither emerge from filaments (lower \orange{red/grey} loop) nor from knots (upper \orange{red/grey} loop).}
  \label{fig:10MeV}
\end{figure*}

\begin{figure}
  \centering
  \noindent
  \resizebox{\columnwidth}{!}{
  \includegraphics{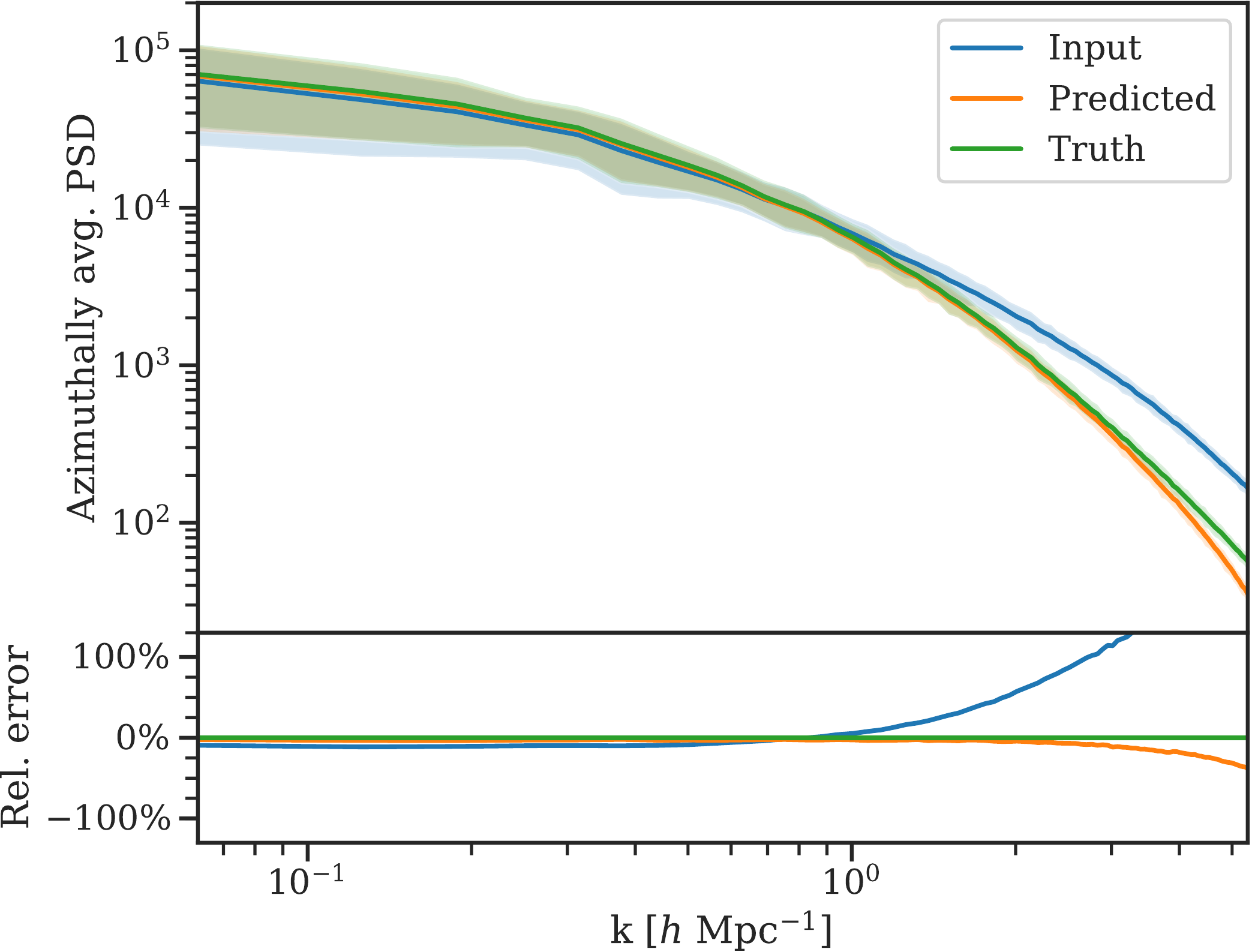}
  }
  \caption{\emph{No DMAF $\to$ DMAF}: azimuthally averaged power spectral density (PSD) of the \orange{test images. The solid lines depict the sample mean and the shaded areas show the $2\sigma$-region}. The relative error towards the ground truth images is plotted in the lower \orange{panel}. On large scales (i.e. small $k$), the PSDs of input, ground truth, and predicted images agree well with each other. On smaller scales, the PSD of the input images declines more slowly due to the finer substructure as compared to the output images. On even smaller scales consisting of a few pixels, the PSD of the predicted output images starts to drop with respect to the ground truth output.}
  \label{fig:PSD_10MeV}
\end{figure}

\begin{figure}
  \centering
  \noindent
  \resizebox{\columnwidth}{!}{
  \includegraphics{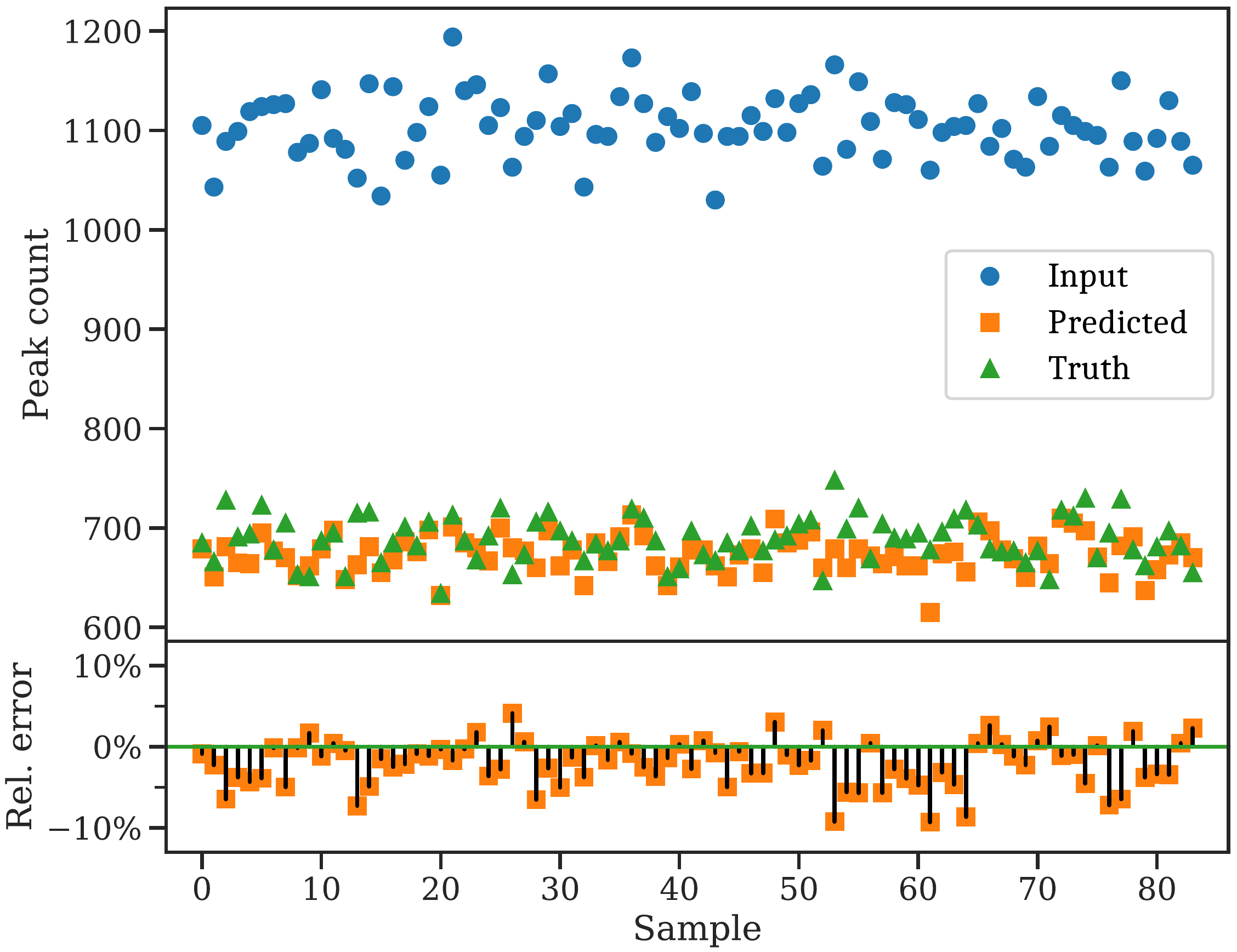}
  }
  \caption{\emph{No DMAF $\to$ DMAF}: gas density peak counts for all $84$ images from the test set after applying a Gaussian filter with $\sigma = 1$ pixel to eliminate pixel-to-pixel noise. DMAF reduces the average number of peaks in slices from cosmological simulations by $38$ per cent, which is well matched by the cGAN predictions that contain on average $2$ per cent less peaks than the ground truth slices with DMAF. The relative error with respect to the ground truth output (see lower panel) is less than $10$ per cent for all samples. }
  \label{fig:peak_count_10MeV}
\end{figure}
Figure \ref{fig:10MeV} shows the input (without DMAF), the prediction by the neural network, and the ground truth (with DMAF) for \orange{a gas density slice} from the test set which the neural network has not seen during training. It is apparent that the amount of substructure in the predicted density \orange{distribution} is reduced as compared to the input and closely resembles the ground truth. Only knots and filaments with sufficient density contrast against the adjacent regions have been retained  whereas finer substructures have been washed out\footnote{In fact, two-dimensional filaments and knots may be cuts through three-dimensional sheets and filaments, respectively, but we use the former notions for intuition.} (see for example the regions highlighted in \orange{black}). Voids appear emptier due to the lack of substructure. The high energy from DMAF has caused some knots to disperse and filaments to fork into bubble-shaped structures (see the regions encompassed in \orange{red/grey}). Evidently, to predict where this happens turns out to be too difficult for the neural network. It is interesting to note that the cGAN has memorised the bubbles in images on which it was trained and correctly reproduces them, but fails to extrapolate the bubble creation to new images. \orange{Note that for more realistic dark matter candidates in a higher mass range, we do not observe these bubbles in our simulations; moreover, they could possibly be diminished by including radiative cooling physics. None the less, bubble-like structures occur in other real scenarios, such as the Fermi bubbles extending north and south from the Galactic Centre, whose origin may be outflows caused by previous accretion onto Milky Way's supermassive black hole or by a past starburst phase of the Milky Way \citep{Su2010}.} Altogether, the visual impression of the gas density slices produced by the neural network is very similar to their original counterparts, save the formation of bubbles. 
\par In order to quantitatively assess the predicted density images, we compare the power spectral densities of the images with each other. Let $\mathbf{I} = (I_{mn}) \in [0, 1]^{M \times M}$ be the intensity value of each pixel. Note that we carry out the analysis for the images, where the intensity corresponds to the \orange{normalised} decimal logarithm of the gas density. The two-dimensional discrete Fourier transform $\mathcal{F}(\mathbf{I}) = (\mathcal{F}_{kl}(\mathbf{I})) \in \mathbb{C}^{M \times M}$ is obtained by computing the matrix entries
\begin{equation}
    \mathcal{F}_{kl}(\mathbf{I}) = \sum_{m=0}^{M-1} \sum_{n=0}^{M-1} I_{mn}\exp\Bigg[-2\upi i \left({\frac{mk + nl}{M}}\right)\Bigg],
\end{equation}
for $k, l \in \{0, \ldots, M-1\}$. From this, we calculate the azimuthally averaged power spectral density of the image as
\begin{equation}
    \text{PSD}_\mathbf{I}(r) = \langle |\mathcal{F}(\mathbf{I})|^2 \rangle_{\text{az}},
\end{equation}
where $\langle \cdot \rangle_{\text{az}}$ denotes an average over the azimuthal angle. Thus, the resulting quantity depends on the radial coordinate $r$ only. 
\par Figure \ref{fig:PSD_10MeV} shows the power spectral density as a function of wave number $k$ for the \orange{density slices from the test set}. The wave number is related to the wavelength $\lambda$ via $k = 2 \upi \lambda^{-1}$. For wave numbers $k > 1 \ {h} \ \text{Mpc}^{-1}$, the washed out substructure due to DMAF is reflected in the drastically reduced spectral power as compared to the input images. The power spectral density of the predicted images is consistent with the ground truth down to scales of $k \sim 2 - 3 \ {h} \ \text{Mpc}^{-1}$; then, the predicted images start to lose power and the neural network begins to introduce a slight blur not observed in the real images. As reported by \citet{Rodriguez2018}, the quality of the power spectrum of GAN-generated images is sensitive to the simulation box size and is expected to improve for larger boxes that give rise to more homogeneous density slices.    
\par A simple but illustrative criterion for evaluating the quality of the generated density slices is peak counts, which has been used in the context of weak lensing (e.g. \citealt{Lin2016, Fluri2018}). A peak is simply defined as a local intensity maximum. As the DMAF smooths the gas distribution, one expects a reduction in the number of peaks, which should be reproduced by the \orange{predictions} from the neural network. In order to exclude very high-frequent noise at the pixel level, we convolve the images with a Gaussian kernel with standard deviation $\sigma = 1 \ \text{pixel}$ before counting the local maxima\footnote{\orange{Taking broader kernels with standard deviations of $2$ and $3$ pixels reduces the total peak counts but leads to similar relative errors between then samples generated by the cGAN and the $N$-body simulation.}}. Pixels at the domain boundary are not taken into account. In Figure \ref{fig:peak_count_10MeV}, the peak counts for all $84$ gas density slices from the test set are shown. Indeed, the peak counts without DMAF exceed the ones with DMAF by more than $60$ per cent for the strong annihilation considered herein. This is well reproduced by the neural network: less than half of the predicted peak counts are off by more than $2.5$ per cent, \orange{roughly} $15$ per cent deviate \orange{by} more than $5$ per cent, and the \orange{maximum relative} error in our test image set amounts to less than $10$ per cent. Consistent with the slightly too low power of the predicted images on very small scales, the average number of peaks is about $2$ per cent smaller compared with the ground truth. 
\section{Inpainting filaments: the reverse problem}
\label{sec:reverse}
\begin{figure*}
  \centering
  \noindent
  \resizebox{\textwidth}{!}{
  \includegraphics{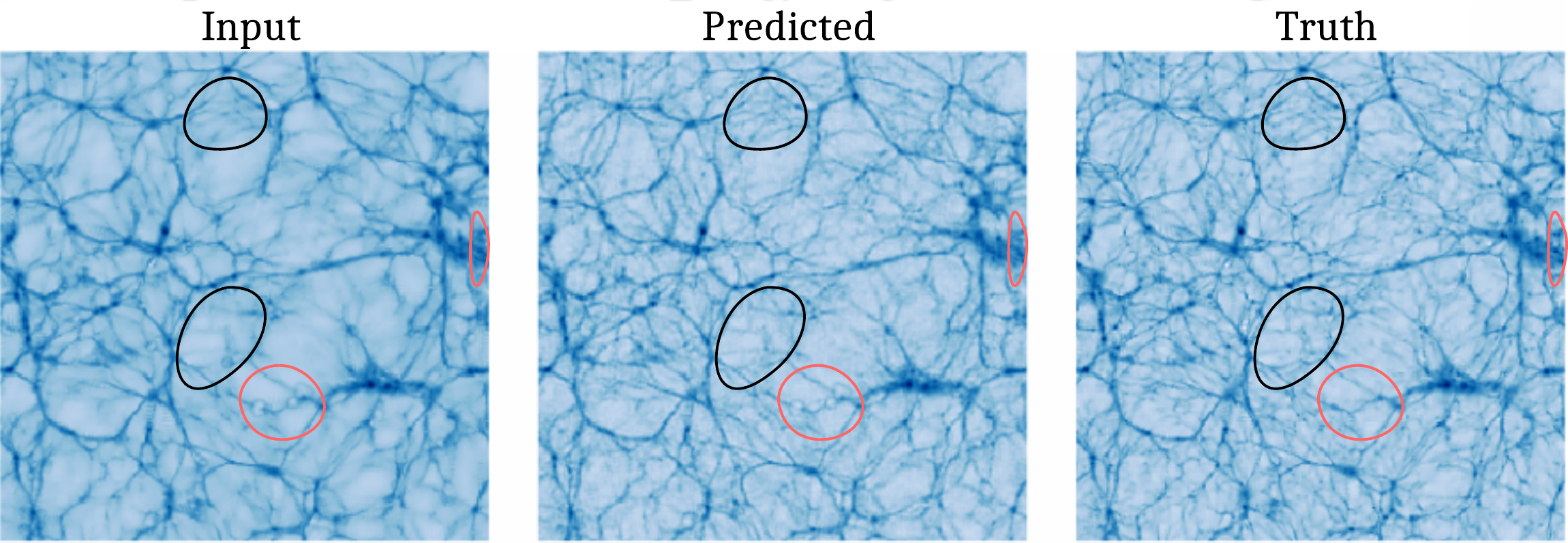}
  }
  \caption{\emph{DMAF $\to$ no DMAF}: predicted gas density distribution for the \orange{image} in Figure \ref{fig:10MeV}. Now, the direction of inference is reversed. The cGAN has added substructure to the smooth input image and drawn filaments based on faint overdensities in the input images (see the areas marked with \orange{black} loops). Bubbles in the large \orange{red/grey} loop have not been erased by the cGAN and individual peaks arising in the ground truth output in the small \orange{red/grey} loop on the right have not been resolved.}
  \label{fig:10MeV_reverse}
\end{figure*}

\begin{figure}
  \centering
  \noindent
  \resizebox{\columnwidth}{!}{
  \includegraphics{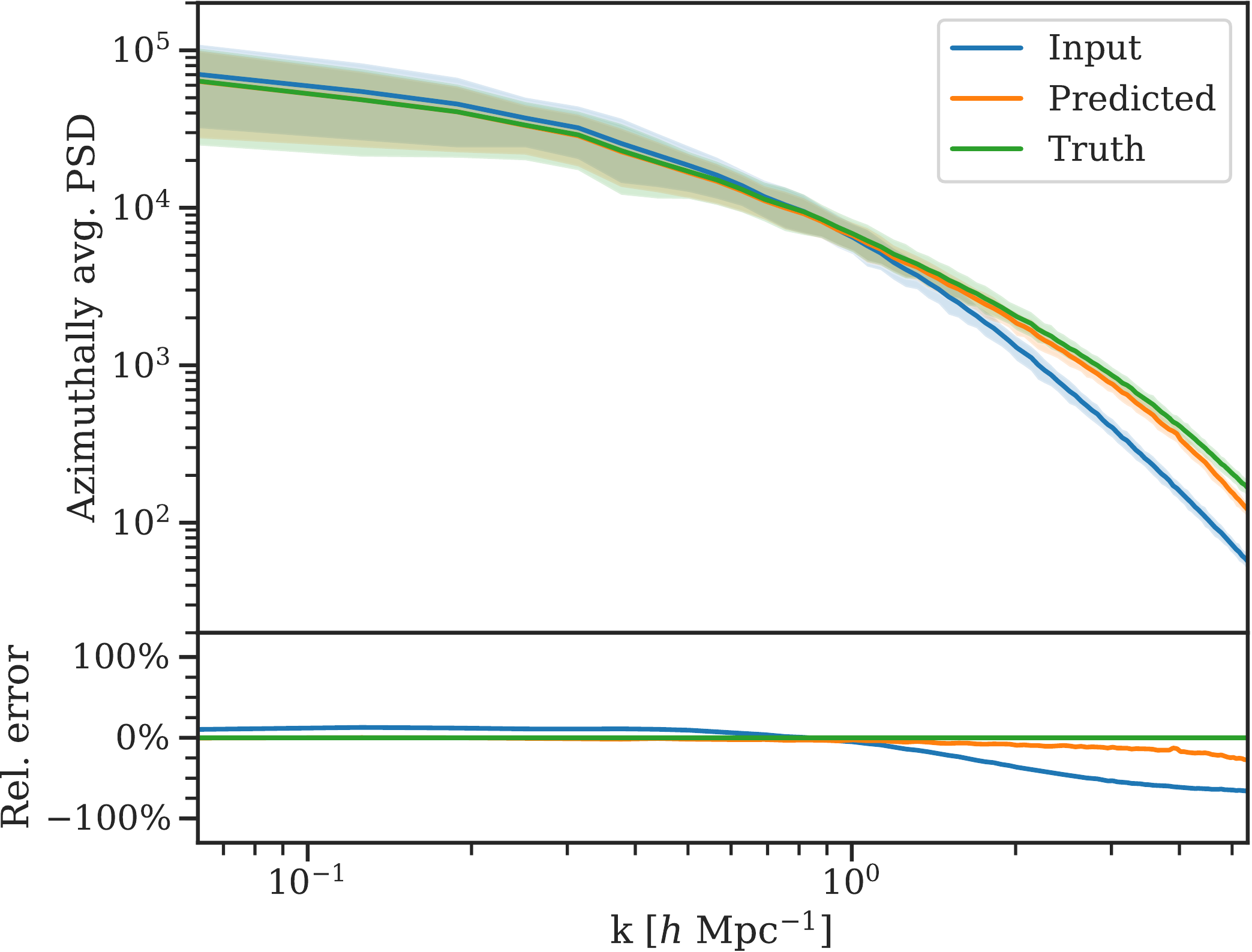}
  }
  \caption{\emph{DMAF $\to$ no DMAF}: azimuthally averaged power spectral density (PSD) of \orange{the test images (sample means and $2\sigma$-regions)}. Note that input and ground truth are now flipped. The relative error towards the ground truth images (without DMAF) is plotted in the lower \orange{panels}. Again, the cGAN-generated images lose power on small scales as compared to the ground truth images.}
  \label{fig:PSD_10MeV_reverse}
\end{figure}

\begin{figure}
  \centering
  \noindent
  \resizebox{\columnwidth}{!}{
  \includegraphics{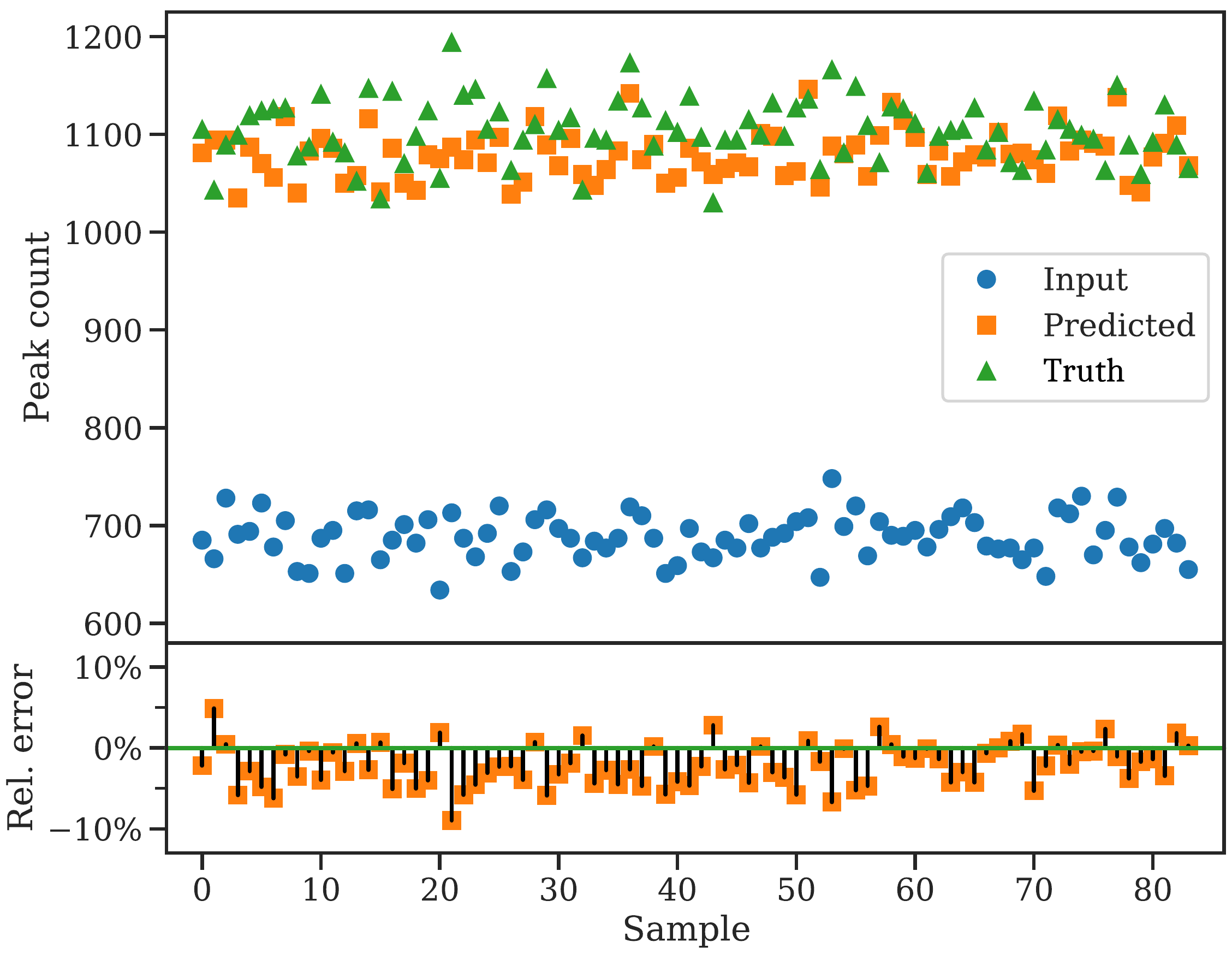}
  }
  \caption{\emph{DMAF $\to$ no DMAF}: gas density peak counts for all $84$ images from the test set as in Figure \ref{fig:peak_count_10MeV} for the inverse direction of inference. Now, the cGAN creates additional substructure, increasing the total peak count. Also here, the cGAN-generated samples contain on average $2$ per cent less peaks than the corresponding ground truth samples without DMAF.}
  \label{fig:peak_count_reverse}
\end{figure}

In the above example, the task of the neural network is to identify structures in the original image which will smear or vanish in the presence of DMAF. Put differently, a more complex density distribution is mapped to a less complex one by the neural network. It is interesting to examine whether the reverse can be achieved, that is whether the neural network can be trained to predict the gas density distribution \emph{without DMAF} given a slice originating from a simulation \emph{with DMAF}. For this, substructure that is not present in the input image (with DMAF) needs to be generated in such a way that the conditioned distribution of predicted output images, given the corresponding input images, is the same as for the ground truth images (without DMAF).
\par We use the same network architecture and training set-up as described above and simply swap input and output images during the training. The predicted output for the first image from the test set is shown in Figure \ref{fig:10MeV_reverse} (cf. \orange{Figure} \ref{fig:10MeV} for the flipped case). The neural network is successful in sharpening filamentary structures and increasing the contrast between filaments and surrounding voids (see e.g. the regions circumscribed by \orange{black} lines). Just as for the \emph{creation} of bubbles, the neural network does not always succeed in \emph{eliminating} bubbles (see the lower region highlighted in \orange{red/grey}). However, the neural network does manage to erase some of the bubbles, as for example in the lower right portion of the lower \orange{black} loop. Despite this and the fact that some of the predicted galaxy clusters lack contrast and separation between the individual peaks (e.g. in the region on the right marked in \orange{red/grey}), the neural network performs decently at recreating a visual impression of the cosmic gas web that resembles \orange{its counterpart from} a cosmological simulation.
\par The power spectral density (Figure \ref{fig:PSD_10MeV_reverse}) and peak counts (Figure \ref{fig:peak_count_reverse}) show the same qualitative behaviour as seen for the inference no DMAF $\to$ DMAF: the power spectral density of the generated samples is correctly reproduced on large scales but decreases too rapidly on small scales. The error in peak counts amounts to less than $10$ per cent for all samples and on average, the cGAN underestimates the number of peaks by $2$ per cent. Hence, we conclude that cGANs can also be of use in situations where a change in physics leads to more complicated distributions of physical quantities. 

\section{Conclusions and outlook}
\label{sec:conclusions}
We have shown that cGANs are able to predict alterations in cosmological simulations induced by a change in physics to a high degree. The smoothing of the gas density field caused by DMAF is accurately reproduced and the statistics of the output images closely resemble those of the images drawn from cosmological simulations with DMAF; although the power spectral density differs on small scales. Other possible applications of this framework range from emulating warm dark matter density fields from cold dark matter simulations over accounting for baryonic feedback (see \citealt{Troster2019} for the related problem of predicting gas pressure fields from the dark matter density distribution using GANs) to varying cosmological parameters. Quantities defined on a three-dimensional grid can be treated within the same framework. In addition, multiple quantities can be integrated as separate image channels (for instance dark matter density, gas density, and temperature), which enables the cGAN to infer correlations between the quantities. An appealing extension to this work consists in feeding parameters (such as the strength of DMAF in the example considered herein) to the neural network and to investigate whether the neural network is able to extrapolate to new parameter ranges as reported by \citet{He2019} for predicting the non-linear evolution of displacement fields. For bidirectional inference (for example DMAF $\leftrightarrow$ no DMAF), CycleGANs \citep{Zhu2017} might be an attractive extension to the cGAN architecture.
\par In the years to come, leveraging the synergy between powerful but expensive methods such as cosmological simulations and fast approximations using deep learning has the potential to exhaust parameter spaces more thoroughly and to augment simulations \emph{a posteriori} with additional physics. 

\section*{Acknowledgements}
The authors acknowledge the National Computational Infrastructure (NCI), which is supported by the Australian Government, for providing services and computational resources on the supercomputer Raijin that have contributed to the research results reported within this paper. This research made use of the Argus Virtual Research Desktop environment funded by the University of Sydney. F. L. is supported by the University of Sydney International Scholarship (USydIS).









\bsp	
\label{lastpage}
\end{document}